\documentclass[%
 reprint,
 superscriptaddress,
 amsmath,amssymb,
 aps,
 prb,
]{revtex4-1}

\usepackage{graphicx}
\usepackage{bm}

\usepackage{array}
	\newcolumntype{L}[1]{>{\raggedright\arraybackslash}p{#1}}
    \newcolumntype{R}[1]{>{\raggedleft\arraybackslash}p{#1}}
\usepackage{siunitx}
\AtBeginDocument{\usepackage{booktabs}}
\usepackage[flushleft]{threeparttable}    
\usepackage{color}

\DeclareMathOperator{\Char}{char}

\newcommand{\mr}[2]{\begin{tabular}{c}#1\\#2\end{tabular}}

\begin{document}

\preprint{APS/123-QED}

\title{A general procedure for accurate defect excitation energies from DFT-1/2 band structures: The case of NV$^-$ center in diamond}

\author{Bruno Lucatto}
 \email{brunolucatto@gmail.com; gmsn@ita.br}
 \affiliation{Grupo de Materiais Semicondutores e Nanotecnologia (GMSN), Technological Institute of Aeronautics (ITA), 12228-900 S\~ao Jos\'e dos Campos, SP, Brazil}
\author{Lucy V. C. Assali}
 \affiliation{Institute of Physics, University de S\~ao Paulo (USP), 05315-970 S\~ao Paulo, SP, Brazil}
\author{Ronaldo Rodrigues Pela}
 \affiliation{Grupo de Materiais Semicondutores e Nanotecnologia (GMSN), Technological Institute of Aeronautics (ITA), 12228-900 S\~ao Jos\'e dos Campos, SP, Brazil}
\author{Marcelo Marques}
 \affiliation{Grupo de Materiais Semicondutores e Nanotecnologia (GMSN), Technological Institute of Aeronautics (ITA), 12228-900 S\~ao Jos\'e dos Campos, SP, Brazil}
\author{Lara K. Teles}
 \affiliation{Grupo de Materiais Semicondutores e Nanotecnologia (GMSN), Technological Institute of Aeronautics (ITA), 12228-900 S\~ao Jos\'e dos Campos, SP, Brazil}

\date{\today}

\begin{abstract}
\noindent

A major challenge in creating a quantum computer is to find a quantum system that can be used to implement the qubits.
For this purpose, deep centers are prominent candidates, and \emph{ab initio} calculations are one of the most important tools to theoretically study their properties.
However, these calculations are highly involved, due to the large supercell needed, and the computational cost can be even larger when one goes beyond the Kohn-Sham scheme to correct the band gap problem and achieve good accuracy.
In this work, we present a method that overcomes these problems and provides the optical transition energies as a difference of Kohn-Sham eigenvalues; and even more, provides a complete and accurate band structure of the defect in the semiconductor.
Despite the original motivations, the presented methodology is a general procedure, which can be used to systematically study the optical transitions between localized levels within the band gap of any system.
The method is an extension of the low-cost and parameter-free DFT-1/2 approximate quasi-particle correction, and allows it to be applied in the study of complex defects.
As a benchmark, we apply the method to the NV$^-$ center in diamond.
The agreement with experiments is remarkable, with an accuracy of 0.1~eV.
The band structure agrees with the expected qualitative features of this system, and thus provides a good intuitive physical picture by itself.

\end{abstract}

\maketitle

\pagestyle{myheadings}
\markboth{BRUNO LUCATTO}{ACCURATE EXCITATION ENERGIES OF THE NV$^-$...}
\thispagestyle{empty}

\section{Introduction}

One of the most exciting engineering problems of current days is to develop a quantum computer.
A quantum computer is a computation system that makes direct use of quantum phenomena, such as entanglement and superposition, to perform operations on data.
Their fundamental building blocks are called qubits, in analogy to the bits present in digital computers.
Quantum computers could enable us to solve complex and time-demanding problems in a much faster way.
This performance difference is not due to an eventual faster clock speed, but due to the different kind of operations that quantum computers will be able to perform with the data stored in qubits. \cite{Struck2016}

A major challenge in creating a quantum computer is to find a quantum system that could be used to implement the qubits. Most systems interact strongly with their surroundings, causing decoherence and consequently loss of information.
In this scenario, deep centers are prominent. \cite{Morton2011}
They are point defects in a semiconductor or insulating crystal that bind electrons to a localized region.
Consequently, most characteristics of their electronic states resembles the ones of single atoms or molecules.
Additionally, deep centers  are fixed in space by the surrounding crystal, in contrast to some other proposals that require additional systems, like the magneto-optical traps for ultracold atoms.

A deep center in diamond, known as negatively charged nitrogen-vacancy center (NV$^-$ center), \cite{Doherty2013} has been strongly considered for such applications, since it has many desirable characteristics: its spin can be optically polarized, manipulated with microwaves, optically measured in an on-demand fashion at the single defect level, and also have a huge coherence time, achieving the order of milliseconds. \cite{Koehl2015}

Although NV$^-$ centers in diamond have all the desirable characteristics for a qubit, diamond as a host is not ideal, since it makes device construction and design with current technology really challenging due to its high mechanical resistance and small chemical reactivity. \cite{Koehl2015}
Therefore, it is desirable to find deep centers with NV-like properties in semiconductor hosts that are more technologically mature, and a systematic search has been initiated. \cite{Gordon2013}

To predict theoretically whether a system is suitable to implement a qubit, first principles calculations based on density functional theory (DFT) are often used. 
However, in order to achieve enough accuracy, the calculations are usually highly involved because one needs to use a large supercell and one must go beyond the Kohn-Sham scheme to better describe it, due to the underestimation of the quasi-particle band gap in standard DFT calculations.
Hence, it is desirable to develop a fast method that can be employed on this systematic search for NV$^-$ like systems.
A good choice is the LDA-1/2 (\textit{LDA minus half}), developed by Ferreira, Marques and Teles, \cite{Ferreira2008,Ferreira2011} which introduces approximate quasi-particle corrections and has accurate predictions, while keeping the same computational cost of the standard LDA approach.
The method also works with the more modern GGA functionals, in which case it is called GGA-1/2.
Hence, to avoid unnecessary particularization, we call it DFT-1/2.

In this paper, we extend the DFT-1/2 method to calculate the optical transition energies between the defect levels.
We demonstrate that it is possible to determine the optical transition energies directly from the Kohn-Sham band structure, against the usual procedure of taking the difference between two total-energy calculations.
Our proposal is benchmarked by applying it to the NV$^-$ center in diamond, whose transition energies have already been experimentally determined.
The method allows for a systematic search for deep centers, since it has as a final result a complete band structure which we can use to accurately analyze the system both qualitative and quantitatively.
Moreover, despite the specific nature of its initial motivation, this method is a general procedure, and can be used to study any system with optical transition energies between levels within the band gap.

\section{Retrospect for the NV$^-$ center in diamond}\label{s2}

Point defects are usually stable in different charge configurations, depending on the Fermi level position.
The NV center has two different charge configurations, NV$^0$ and NV$^-$, and only the latter has the desired properties to be used as a qubit. \cite{Doherty2013}
Fortunately, nitrogen doping places the Fermi level inside the range where the negatively charged defect is stable. \cite{Weber2010}

\begin{figure}[b]
\centering
\includegraphics[width=0.5\textwidth]{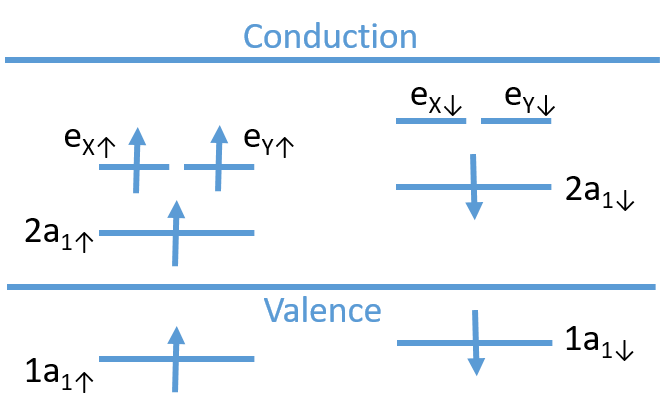}
\caption{(color online) Schematic representation of the defect states and their occupation on the ground state of the NV$^-$ center in diamond. Spin up levels are on the left of the figure, while spin down levels are on the right. There are two distinct levels with $a_1$ geometry, which are labeled in crescent order of energy with a number on the left. The double-degenerate level $e$ is indicated with a broken line.}
\label{simplifiedbands}
\end{figure}

The NV center consists in a substitutional nitrogen atom adjacent to a carbon vacancy, presenting a C$_{3v}$ point symmetry.
An instructive and useful model is to consider the defect as an effective molecule. \cite{Lenef1996,Gomes1985}
This approach, known as ``molecular model for defects'' consists in making symmetry adapted linear combinations of the dangling bonds around the vacancy to construct molecular orbitals.
It has the implicit assumption that the electrons bound to the defect are localized around it and are not ``spilling over'' from the vacancy into the entire crystal. \cite{Coulson1957,Lenef1996}
The states are labeled using the Mulliken symbols.
Lower case letters indicate the single-particle states, while capital letters label the many-body states.



Since each dangling bond from the three carbon atoms surrounding the vacancy contributes with one electron, and the overlapping nitrogen lone pair has other two electrons, we conclude that the neutral NV center would have five electrons and, consequently, the NV$^-$ would have six, which would be accommodated in the defect energy levels.

We note that the defect has states within the band gap that are spin dependent, a consequence of the fact that this defect breaks the spatial inversion symmetry of the crystal.
The electronic occupation for the ground state of the NV$^-$ center is shown in Fig.~\ref{simplifiedbands}.
It can be obtained by filling the lowest energy states with the corresponding spin state (spin up on the left side, spin down on the right side).
This leaves us with four spin-up and two spin-down electrons, hence the spins in the ground state do not cancel out and we have a total effective spin $S=1$, \emph{i.e.} the ground state is a triplet.
This fact is of central importance in the application of the NV$^-$ center as a qubit, since it is the spin that is used to store the quantum information.

\begin{figure}[b]
\centering
\includegraphics[width=0.5\textwidth]{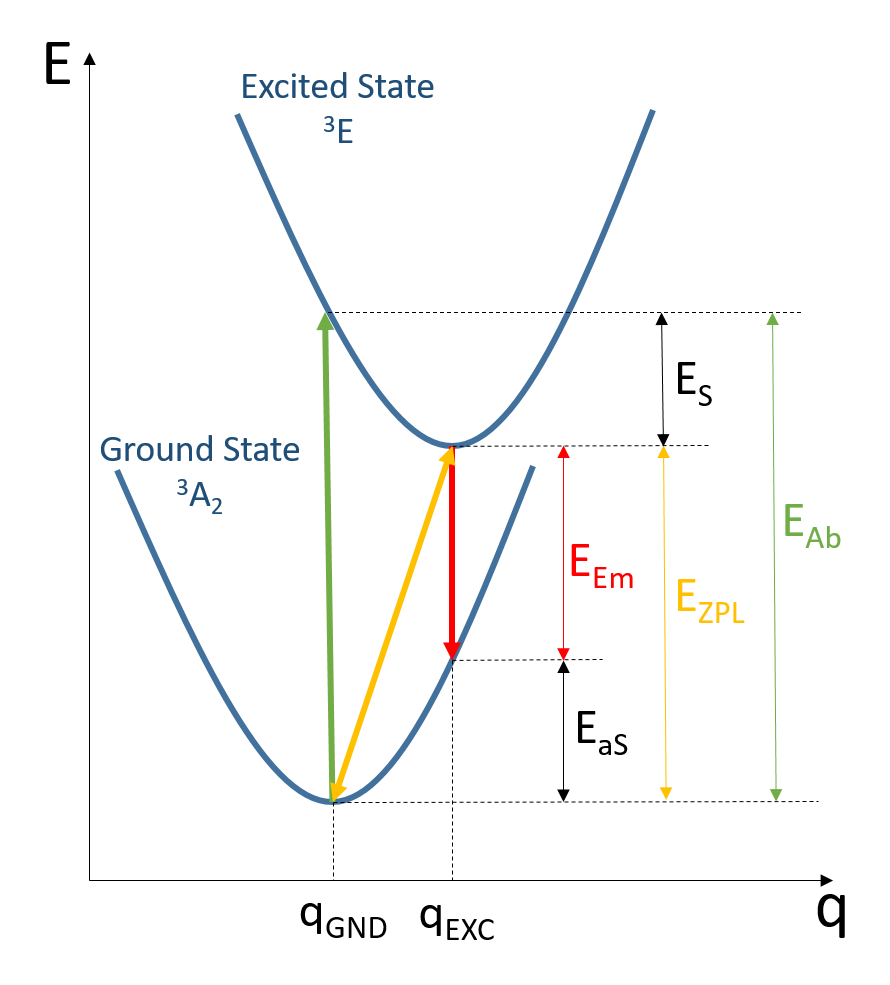}
\caption{(color online) Schematic configuration coordinate diagram for the NV$^-$ center. The curves represent the energy of the defect as a function of the displacement of the atoms, measured by a generalized coordinate q, for both its many-body electronic ground and excited states. The minima of these curves correspond to the relaxed geometry of each case. The vertical transitions (green and red arrows) correspond to the peaks in the optical absorption and emission curves, respectively. The transition between the minima is called the zero-phonon line (ZPL), and is the same for both the emission and absorption. The Stokes (E$_S$) and anti-Stokes (E$_{aS}$) shift energies are also indicated.}
\label{energias}
\end{figure}

When electromagnetic radiation of 2.18~eV (569~nm, green light) is applied, \cite{Davies1976} as depicted in Fig~\ref{energias}, there is a resonant excitation to the first excited state.
This can be understood in light of  Fig.~\ref{simplifiedbands} as promoting the spin down electron in the state $2a_{1\downarrow}$ to one of the excited states $e_{x\downarrow}$ or $e_{y\downarrow}$.
Note that this is the first possible optical excitation of the system, since changes in spin are forbidden at the first order.
Another important fact concerning this transition is that it is possible to excite the system without exciting electrons neither from the valence band to the defect nor from the defect to the conduction band, because the defect levels are far apart from the Valence Band Maximum (VBM) and the Conduction Band Minimum (CBM).
If the levels were shallower, it would be possible to excite electrons into or outside the defect levels, what would compromise its operation as a qubit.

Considering the multi-electronic system, both the ground state and the excited state are triplets.
However, the excited state transforms as the E symmetry representation, in contrast with the ground state, which transforms as A$_2$.
This difference in symmetry of the wave function impacts the geometry of the defect.
After a transition, the structure relaxes to the new equilibrium geometry, and since the movement of the ions is orders of magnitude slower than the electronic excitation, it is a good approximation to consider that the absorption and emission corresponds to vertical transitions (Fig~\ref{energias}).
The photon emission then occurs in the geometry of the excited state, and the difference in energy between the excited state and the ground state in this configuration is 1.76~eV (704~nm, red light). \cite{Davies1976}
The difference in energy between the excited and ground states in their respective relaxed geometries is called the Zero Phonon Line (ZPL), and is also indicated in Fig~\ref{energias}.
This considerable difference in the wavelength enables to easily separate the photons of the pumping laser from the photons emitted by the center by using a dichroic mirror (which reflects one wavelength and transmits the other), as in confocal microscopy.

It is important to mention that the DFT approach and the defect molecular model are complementary to each other.
These two theoretical methods have their complementary strengths and weaknesses, and only their combined application can give us a good picture of the observed phenomena. \cite{Doherty2013}

\section{The DFT-1/2 local correction to defect levels}

The most important tools to support the search for suitable deep centers for quantum computing applications are the \emph{ab initio} computational techniques based on DFT, since they allow us to determine macroscopic properties based only on the system's atomic composition and approximate geometry.
Many attempts have been made to find such defects to find such kind of defects in several semiconductors, as in some silicon-carbide polytypes, \cite{Koehl2015,Weber2010} wurtzite aluminum nitride \cite{Tu2013} and zinc-blende gallium nitride. \cite{Wang2012}

Two major concerns can make first-principle calculations of defects a difficult task.
First, the usually employed periodic boundary conditions to study solids require a large supercell to minimize the interaction between the defect and its images.
Second, the Kohn-Sham band gap is underestimated when compared to experiments, \cite{Sham1985,Perdew1983} which also impairs reliable calculations of defect levels above the valence band. \cite{Rinke2009}
The method which correct Kohn-Sham eigenvalues, such as hybrid functionals \cite{Heyd2003,Heyd2006} and the \emph{GW} approach, \cite{Hybertsen1985} usually raise the computational cost. \cite{Pela2015}
The DFT-1/2 method is a good alternative due to its nice accuracy and low computational cost. 


LDA-1/2 and GGA-1/2 have already been successfully used to study point defects \cite{Matusalem2013,Matusalem2014} by applying a formalism developed by Rinke \emph{et al.} \cite{Rinke2009}
In these cases, the interest was to study the defect formation and transition energies, both quantities related to the electronic ground state in several charge states, such that what was changing was the number of electrons binded to the defect as a function of the Fermi level position.
In the present case, the charge of the defect is always the same and our interest is to study the energies associated to the optical excitation of an electron between intra-defect energy levels.



The DFT-1/2 method generalizes the Slater's transition state technique for solids, introducing approximate quasi-particle corrections which lead to accurate band gap calculations.
The details of the method are given in Refs.~\citenum{Ferreira2008,Ferreira2011}.
In this work, we present an overview to contextualize the reader who is not familiar with the method. The approach relies on the Janak's theorem \cite{Janak1978} and on the approximately linear dependence of the Kohn-Sham eigenvalues with its own occupation. 
It is possible to use these two facts to show that, in the case of atoms and molecules, the value of the highest occupied eigenvalue with half-ionization is the system ionization energy with a remarkable agreement with experimental data.\cite{Slater1972}

In semiconductors and insulators, the quasi-particle band gap is defined as the energy difference between the ionization energy and the electronic affinity.
Thus, this scheme allows us to compute the band gap as the difference between the Kohn-Sham eigenvalues, by introducing a half-hole on the VBM and a half-electron on the CBM.

\begin{figure}[t]
\centering
\includegraphics[width=0.5\textwidth]{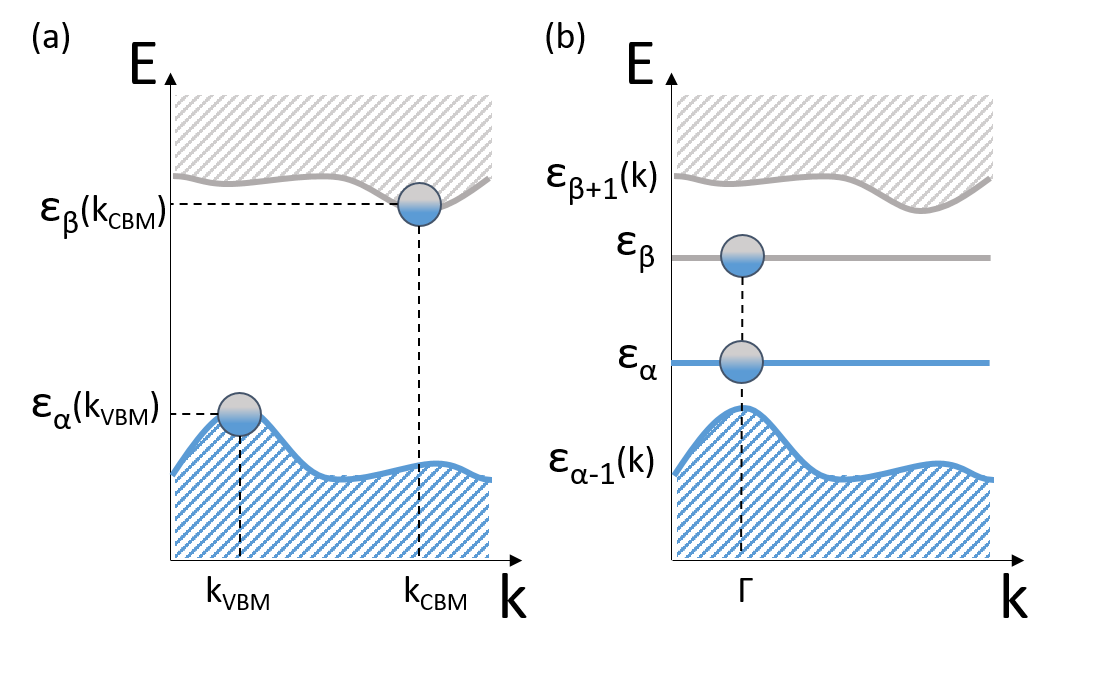}
\caption{(color online) (a) Schematic representation of a Kohn-Sham band structure with the Valence Band Maximum and Conduction Band Minimum with half-occupation, as considered on the DFT-1/2 method; and (b) Extension of the DFT-1/2 scheme for defect levels within the band gap.}
\label{KSdiagrams}
\end{figure}

Since Bloch states are delocalized, they do not accurately describe neither the hole on the valence band nor the electron on the conduction band. \cite{Ferreira2011}
Therefore, instead of changing the occupations of the levels, this contribution in energy is added to the potential of the atoms itself.
It is assumed that this potential has the same format of the atomic self-energy potential $V_S$, which can be simply computed as the difference between the neutral atomic potential and the half-ionized atomic potential. \cite{Ferreira2008}
Considering that the localized hole state will be close to the VBM and the localized electron state will be close to the CBM, we must find which atomic orbitals contribute to each of these levels and in what proportion (the orbital \emph{character} of the levels).
This is quantified by the projection of these Kohn-Sham orbitals onto the atomic orbitals.
A schematic representation of this scenario is given on Fig.~\ref{KSdiagrams}(a).

In what follows, we describe an extension of the method for defect levels, which resembles in many aspects the scheme for the bulk (Fig.~\ref{KSdiagrams}(b)).
Due to the increased complexity of the orbital character of the levels, a more precise notation is necessary.
Indeed, this is also a formalization of some ideas that already have been introduced in recent publications. \cite{Freitas2016,Freitas2016a,Ataide2017}

We must add the potential that corresponds to the removal of half-electron from the occupied level (labeled $\alpha$).
In the usual and simple cases for the bulk, the Kohn-Sham state $\psi_\alpha(\rm{k_{VBM}})$ is composed only of the valence level $p$ orbital of the ion.
However, in the case of the defect, we can have a set of atoms contributing to this level, in which case we must remove a smaller fraction of electron from each of them, proportionally to their contribution.
Hence, for each atomic orbital $\phi$ of each atom $X$, we subtract a fraction $\xi_{X,\phi}$ of an electron given by
\begin{equation}
	\xi_{X\phi} = \Char_{X\phi}\big[\psi_\alpha(\Gamma)\big]\times\frac{1}{2},
\end{equation}
where $\Char_{X\phi}[\psi(k)]$ corresponds to the proportion of the atomic orbital $\phi$ of atom $X$ to the orbital character of the Kohn-Sham state $\psi$ at point $k$.
Similarly, we must add the potential that corresponds to the addition of half-electron to the unoccupied level (labeled $\beta$).
The fraction $\zeta_{X\phi}$ to be added to the orbital $\phi$ of atom $X$ is given by:
\begin{equation}
    \zeta_{X\phi} = \Char_{X\phi}\big[\psi_\beta(\Gamma)\big]\times\frac{1}{2}.
\end{equation}

The projection on atomic orbitals is usually a standard output of DFT codes, and is computed as the projection of the wave functions onto spherical harmonics within spheres of an atomic species-dependent radius around each ion.
Considering the fact that some small contributions of atoms far from the defect are going to be neglected, it is important to normalize the orbital characters of the considered atoms with respect to their sum, ensuring that
\begin{equation}
	\sum_{X\phi}\xi_{X\phi} = \sum_{X\phi}{\zeta_{X\phi}} = \frac{1}{2}.
\end{equation}

The self-energy potentials are considered spherically symmetric, so the dependence on $r$ will be omitted on our notation. We compute the components $V^{X\phi}_S$ of the self-energy potential $V_S$ as
\begin{align}
	V_{S,\alpha}^{X\phi} &= V_{X}\big(f_0\big)-V_{X}\big(f_0-\xi_{X\phi}\big) \\
    V_{S,\beta}^{X\phi} &= -\big[V_{X}\big(f_0\big)-V_{X}\big(f_0-\zeta_{X\phi}\big)\big],
\end{align}
where $f_0$ is the occupation of the orbital $\phi$ of atom $X$ on the ground state, and $V_X(f)$ is the potential of atom $X$ with occupation $f$.
Adding the components, we find
\begin{equation}
	V_S^{X\phi} = V_{X}\big(f_0-\zeta_{X\phi}\big)-V_{X}\big(f_0-\xi_{X\phi}\big).
\end{equation}


Before adding the potentials to the Kohn-Sham potential, we must multiply them by a trimming function $\Theta_{X\phi}(r)$ to avoid the divergence that would arise from the sum of the $1/r$ coulombic tails of these potentials.\cite{Ferreira2008,Ferreira2011}
$\Theta$ is a smooth step-like function, defined as
\begin{equation}
	\Theta(r) =
    \begin{cases}
    	\bigg[1-\Big(\frac{r}{CUT}\Big)^8\bigg]^3 & \text{if $r \le CUT$} \\
        0                                         & \text{if $r > CUT$}
    \end{cases}
\end{equation}
which depends on a parameter called $CUT$.
This parameter have to be determined variationally, by following the same procedure as in the bulk case, i.e., by extremizing the difference between the considered levels. \cite{Ferreira2008,Ferreira2011}
Thus, the trimmed potential is
\begin{align}
	\widehat{V}_{S}^{X\phi} &= \Theta_{X\phi}V_{S}^{X\phi}.
\end{align}

\begin{figure*}[t]
\centering                     
\includegraphics[width=\textwidth]{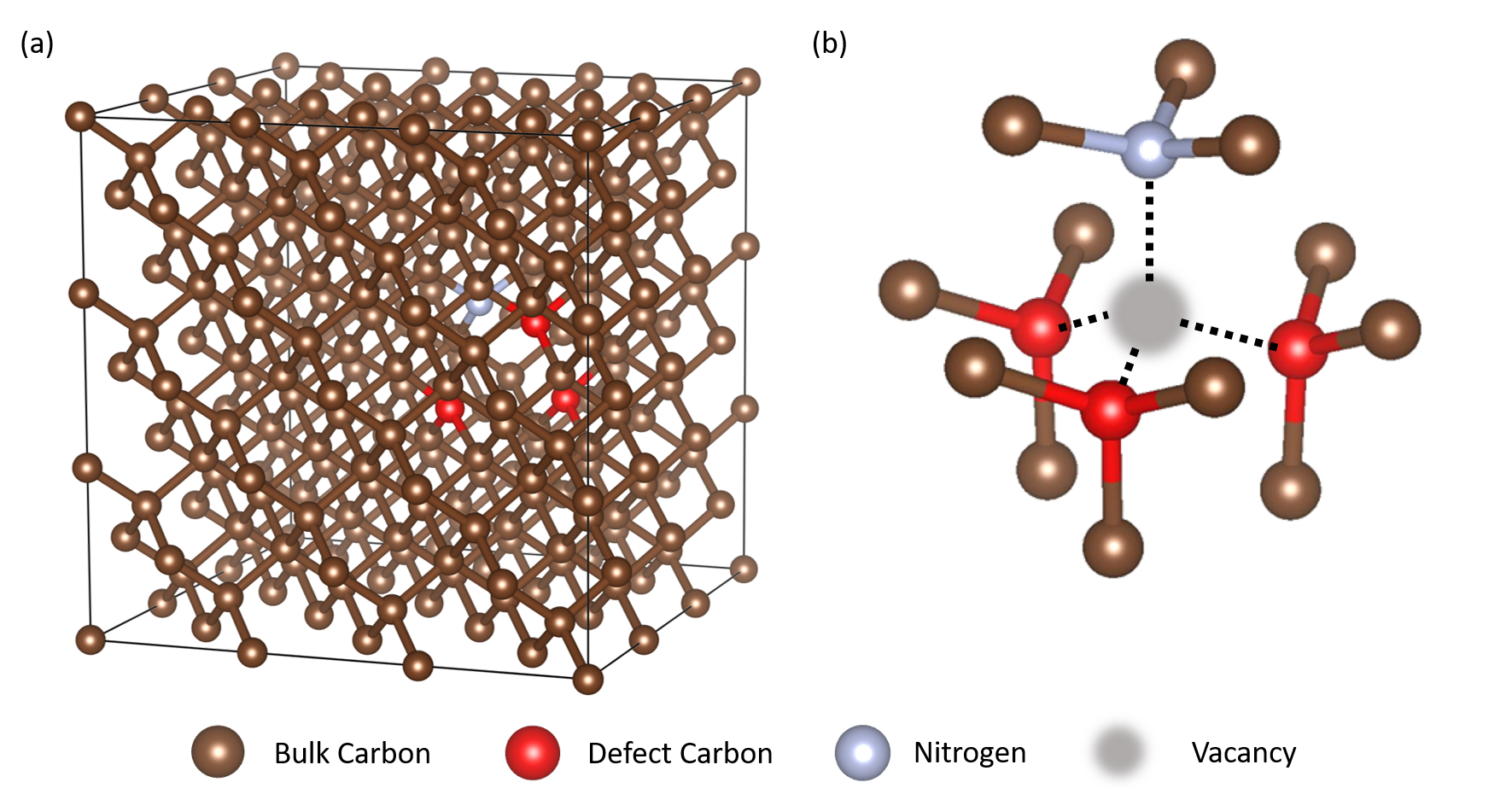}
\caption{(color online) (a) 215-atom supercell used to simulate the defect avoiding spurious interactions among images.
(b) NV$^-$ center and its surrounding atoms, representing the vacancy as a gray shadow.
In both images, Brown, bluish-gray and red circles represent, respectively, the host carbon atoms, the nitrogen atom, and the three carbon atoms neighboring the vacancy.
The images have been produced with help of the VESTA software. \cite{Momma2011}}
\label{supercell}
\end{figure*}

It is common to have situations in which the CUT depends only on the atom. In these cases, it is useful to define 
\begin{align}
	\widehat{V}_{S}^{X} &= \Theta_{X}\sum_\phi{V_{S}^{X\phi}},
\end{align}
and then we would have a single correction to the potential per atomic specie, with a single value of CUT to be determined variationally.

The most noticeable difference between the usual DFT-1/2 and the procedure here introduced is that in the latter exactly half electron is transferred between the defect levels, being divided amongst the atoms which contribute to them.
In solids, the total number of transferred electrons scales with the number of atoms in the cell, since the corrections are applied as if each atom contributed independently to the composition of the VBM and the CBM.


\section{Computational details}

\begin{figure*}[t]
\centering
\includegraphics[width=\textwidth]{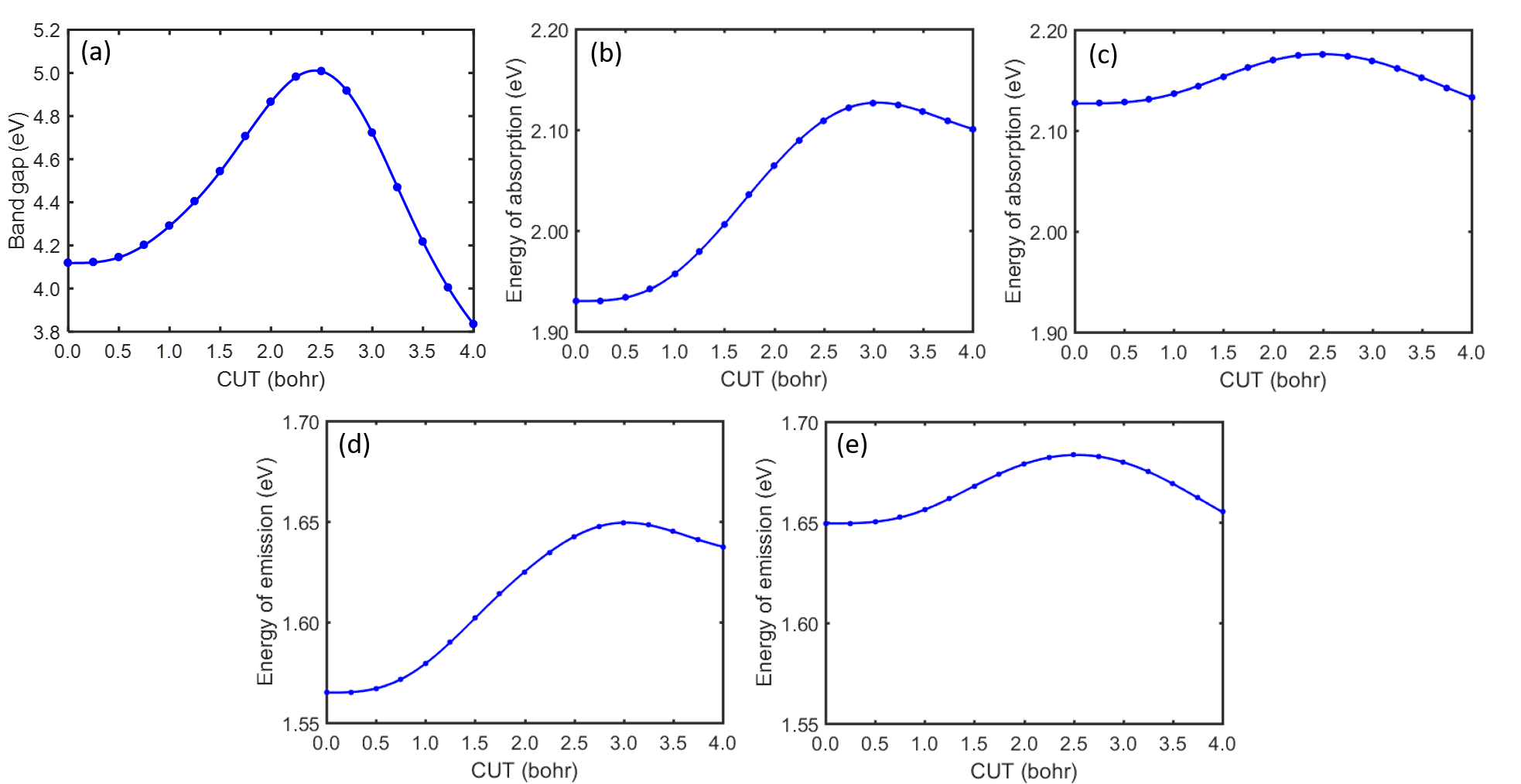}
\caption{(color online) CUT determination for the DFT-1/2 corrections. The corrections have been performed sequentially for each geometry, in the following order: C$_{Bulk}$, N, C$_{Defect}$. (a) Band gap of pure diamond as a function of the CUT for the 0.25 electron removal from the C$_{Bulk}$ atoms' $2p$ orbital; (b) Transition energy on the ground state's geometry as a function of the CUT of the nitrogen atom, with C$_{Bulk}$ corrected; (c) Transition energy on the ground state's geometry as a function of the CUT of the C$_{Defect}$ atoms, with C$_{Bulk}$ and N corrected; (d) Transition energy on the excited state's geometry as a function of the CUT of the nitrogen atom, with C$_{Bulk}$ corrected; (e) Transition energy on the excited state's geometry as a function of the CUT of the C$_{Defect}$ atoms, with C$_{Bulk}$ and N corrected.}
\label{cutsweep}
\end{figure*}

The calculations have been performed within the DFT combined with the Generalized Gradient Approximation of Perdew-Burke-Ernzerhof (GGA-PBE) exchange-correlation potential \cite{Perdew1996} using the Viena Ab-initio Simulation Package (VASP). \cite{Kresse1996,Kresse1996a}
The electronic wave functions have been expanded using the projected augmented wave (PAW) method. \cite{Blochl1994,Kresse1999}

In order to build a good approximation for the supercell structure, first the structure of a single cubic unit cell has been relaxed.
The next step is to replicate it side by side to build a 3x3x3 supercell, with a total of 216 atoms, and a new structural relaxation has been carried out.
Then, the defect has been created by arbitrarily removing one of the atoms of the supercell and replacing one of the carbon atoms neighboring the resulting vacancy by a nitrogen (Fig.~\ref{supercell}).
The number of electrons has been increased by one, since we are interested in the negatively charged NV center.
As the last step before applying the DFT-1/2 corrections, a spin-polarized structural relaxation has been performed for both the electronic ground state and first excited state by setting the corresponding energy levels occupations, to obtain and store the respective resulting atomic positions.
It is noteworthy that, according to  Fig.~\ref{simplifiedbands}, the first excited state corresponds to promote the highest occupied spin-down state ($2a_{1\downarrow}$) to the lowest unoccupied spin-down states ($e_{x\downarrow}$ and $e_{y\downarrow}$), with half electron in each one of the states, to symmetrize the occupation.

In the geometry optimization of the pure cells, \emph{i.e.} cells that do not include the defect, the volume and shape of the cell and all the atoms have been allowed to relax until the magnitude of all forces is smaller than 10$^{-3}$ eV/\AA.
In order to save computational effort and relying on the fact that only the nearest atoms should be affected by the defect, the volume and shape of the cell have been fixed for the relaxation of the supercells with the defect, and the same stopping criteria as before has been used.

Following the Monkhorst-Pack scheme, \cite{Monkhorst1976} the Brillouin zone (BZ) has been sampled by a 19x19x19 grid of k-points for the single cubic cell and by only the gamma point for the supercells.
The plane wave basis set has been considered within a cutoff energy of 530~eV.
The electronic convergence criterion has been that the total (free) energy and the band structure energy change between two steps are both smaller than 10$^{-7}$~eV.
In the simulation of the negatively charged defect, a positive uniform background charge has been added.
The numeric errors of our calculations have been estimated to be smaller than 50~meV.

\section{Applying the local correction to the NV$^-$ center in diamond}\label{localcorrection}

The usual procedure to calculate the optical transition energies of defect levels is to take the difference in total energy between each electronic configuration.
Because this concerns excitations, one needs to go beyond the Kohn-Sham scheme, by carrying out \emph{e.g.} HSE calculations, to avoid the usual band gap problem.\cite{Rinke2009}
Hence, one must go beyond standard DFT, as in HSE calculations, in order to obtain a more accurate result for these energies.
The inconvenient is the increase in the computational cost.
Therefore, it would be of interest to apply the less demanding DFT-1/2 formalism.
However, as implemented, this method does not compute a physically meaningful total energy, but the optical transition energies can be accurately obtained as the difference between their corresponding Kohn-Sham eigenvalues, as demonstrated in appendix~\ref{eigenvalues}.

\begin{table}[t]
    \centering
    \caption{Orbital character of the defect levels and fractions of electron to be removed from and added to each potential (denoted by $\xi$ and $\zeta$, respectively).}
    \label{tablecorrection}
    
    \begin{tabular}{@{}l *{8}{S[table-format=2.1,table-space-text-post=*]}@{}}
		\toprule
        & \multicolumn{4}{c}{Ground State} & \multicolumn{4}{c}{Excited State} \\
        \cmidrule(lr){2-5}\cmidrule(l){6-9}
        {X$\phi$} & {$2a_{1\downarrow}$} & {$\xi_{X\phi}$} & {$e_{x\downarrow}$+$e_{y\downarrow}$} & {$\zeta_{X\phi}$} & {$2a_{1\downarrow}$} & {$\xi_{X\phi}$} & {$e_{x\downarrow}$+$e_{y\downarrow}$} & {$\zeta_{X\phi}$} \\
        \midrule
        C$_{2s}$ & 0.6\%  & 0.00 & 3.1\%  & 0.02 & 0.6\%  & 0.00 & 2.2\%  & 0.01 \\
        C$_{2p}$ & 17.7\% & 0.09 & 30.2\% & 0.15 & 21.7\% & 0.11 & 31.1\% & 0.16 \\
        N$_{2s}$ & 4.6\%  & 0.02 & 0.0\%  & 0.00 & 4.9\%  & 0.02 & 0.0\%  & 0.00 \\
        N$_{2p}$ & 40.7\% & 0.20 & 0.3\%  & 0.00 & 28.1\% & 0.14 & 0.3\%  & 0.00 \\
       \bottomrule
    \end{tabular}
\end{table}

\begin{figure*}
\centering
\includegraphics[width=\textwidth]{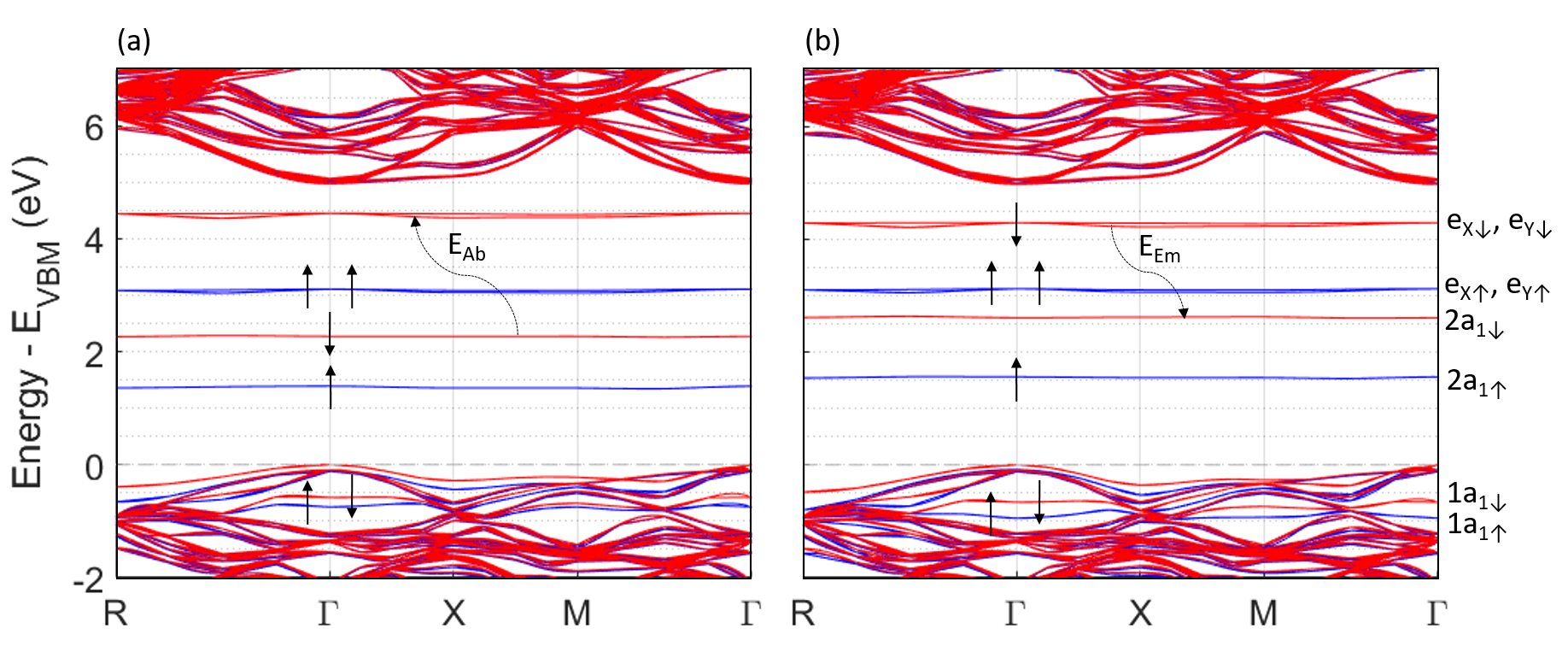}
\caption{(color online) Supercell band structures, around the gap region and along some special high symmetry directions in the cubic BZ, for the structural geometries of the NV- center in the (a) ground state and (b) excited state. The blue and red lines represent, respectively, the spin-up and spin-down energy levels. The occupied states related to the defect are indicated by $\uparrow$ and $\downarrow$ arrows. Both results have been obtained with bulk and local corrections. The direct band gap is a consequence of the supercell band folding.}
\label{bandstructures}
\end{figure*}

We must obtain the difference between the state $2a_{1\downarrow}$ and the double-degenerated state $e_{\downarrow}$ in each geometry, according to Fig.~\ref{simplifiedbands}.
We may write
\begin{align}
	E_{Ab}&=\varepsilon(e_{\downarrow};q_{gnd})-\varepsilon(2a_{1\downarrow};q_{gnd}) \\
    E_{Em}&=\varepsilon(e_{\downarrow};q_{exc})-\varepsilon(2a_{1\downarrow};q_{exc}),
\end{align}
where $\varepsilon(\psi;q)$ corresponds to the eigenvalue of the state $\psi$ as a function of the configuration coordinate $q$, which in the current case correspond to the most stable geometries in each one of the two electronic configurations, as indicated in Fig.~\ref{energias}.
Note that it is only possible to unambiguously define these functions because the position of the eigenvalues are considered independent of the occupation of these levels, as explained in the appendix~\ref{eigenvalues}.
Otherwise, they would be functions of the occupation as well.

With the relaxed geometries for both the ground and excited states, we first separate the atoms in three types: the carbon atoms of the bulk (C$_{Bulk}$), which are responsible for the valence and conduction bands, the nitrogen atom, and the carbon atoms which are the carbon vacancy next-neighbors (C$_{Defect}$), whose dangling bonds contribute to the localized defect levels.
Since these types contribute differently to the band structure, they must be analyzed separately.

For the C$_{Bulk}$ atoms, the same corrected potential as the one used in the diamond unit cell is applied: due to the perfectly covalent bonds between the carbon atoms, the band gap of diamond is corrected by subtracting one quarter of electron from the $2p$ orbital of the C$_{Bulk}$ atoms, as indicated in Ref.~\citenum{Ferreira2008}.
The CUT of 2.5~bohr is determined by maximizing the band gap, which gives a gap of 5.01~eV, as shown in Fig.~\ref{cutsweep}(a).

To apply the local correction to the defect, the character of the levels involved in the first optical excitation, for both geometries, is determined by using the band character obtained by standard PBE calculation.
Accordingly, the percentage of the character contribution is obtained by considering solely the nitrogen atom and the carbon atoms that are the vacancy next-neighbors (C$_{Defect}$).
Table~\ref{tablecorrection} presents the orbital character of the defect levels.

For each orbital, in each geometry, the potential of half electron, weighted by the character of the $2a_{1\downarrow}$ level, must be removed, while the potential of half electron, weighted by the character of the two states $e_{\downarrow}$, must be added.
These results are also displayed in Table~\ref{tablecorrection}.
The CUT parameters for these corrections are determined by maximizing the difference between the levels $e_{\downarrow}$ and $2a_{1\downarrow}$, and we obtain the same values for the excited and ground states, which are CUT=2.50~bohr for C$_{Defect}$ and CUT=3.00~bohr for N.
The curves obtained in this optimization procedure are displayed in Figs.~\ref{cutsweep} (b, c, d and e).

The maximum values obtained on Figs.~\ref{cutsweep} (c and e) correspond, respectively, to the corrected absorption and emission energies.
The values are displayed in Table~\ref{tableresults}, together with other results and experimental data.
With their respective corrected potentials, the electronic structure is calculated for each geometry, and the corrected band structures are displayed in Fig.~\ref{bandstructures}.
The energy differences between the defect levels correspond to the optical transition energies, as indicated by the curved arrows. 
The similarity of these results with the initial and simple picture of the position of the energy levels (Fig.~\ref{simplifiedbands}) is remarkable.
A discussion about Table~\ref{tableresults} and about the band structures shown in Fig.~\ref{bandstructures} is given in Section~\ref{results}.

Besides the vertical transition energies, the ZPL energy is also of experimental interest.
We cannot simply use the difference between the Kohn-Sham eigenvalues of two different geometries to calculate it, since it would not take into account the energy difference due to the displacement of the ions.
Nonetheless, we can indirectly calculate E$_{ZPL}$.
This is possible because the values for the total energy of the two geometries in their electronic ground state are correctly calculated by the standard DFT.
Hence, we can obtain the anti-Stokes shift as
\begin{equation}
	\label{AS} E_{aS}=E(f_{gnd},q_{exc})-E(f_{gnd},q_{gnd}),
\end{equation}
where $E(f,q)$ is the energy as a function of the electronic configuration $f$ and the geometry $q$, and use it together with the vertical transitions to determine the remaining desired energies as
\begin{eqnarray}
    \label{ZPL} E_{ZPL}=E_{Em}+E_{aS}\\
    \label{SS}  E_{S}=E_{Abs}-E_{ZPL},
\end{eqnarray}
as one can readily see from Fig.~\ref{energias}.

Finally, the steps to be followed to apply the DFT-1/2 method for defect levels, introducing Local Corrections, can be summarized:
\begin{enumerate}
	\setlength{\itemsep}{0pt}
	\item Perform the structural relaxation of the unit cell;
	\item Determine the VBM and CBM orbital characters;
	\item Build the supercell and perform a new structural relaxation;
    \item Set and build the defect in the supercell, perform the structural relaxation with the electronic ground state occupancy and determine the orbital character of the selected defect levels;
    \item Perform the structural relaxation with the electronic excited state occupancy and determine the orbital character of the selected defect levels;
    \item Calculate the system total energy in the electronic ground state in the excited state geometry, and determine the anti-Stokes shift (E$_{aS}$) using Eq.~\ref{AS};
    \item Determine the CUT parameter for the bulk atoms by maximizing the band gap;
    \item Determine the CUT parameter for the defect atoms by maximizing the energy difference between the selected defect levels, and determine the energies of the vertical transitions (E$_{Ab}$ and E$_{Em}$); and
    \item Determine the remaining energies (E$_{ZPL}$ and E$_{S}$) using Eq.~\ref{ZPL} and Eq.~\ref{SS}, respectively.
    \item Optional: Calculate the corrected band structures.
\end{enumerate}

\section{Results and Discussion}\label{results}

The diamond band gap value of 5.01 eV obtained with GGA-1/2 approach shows a remarkable improvement over the value of 4.1 eV obtained with standard GGA, when compared to the exprimental value of 5.47 eV.\cite{Wort2008}
Even though the result has a considerably better agreement with the experimental value, it is still slightly underestimated, not as good as the corrections to other materials.\cite{Ferreira2011}
This is due to the fact that the VBM and CBM of diamond's band structure have almost the same orbital character and the usual procedure \cite{Ferreira2008, Ferreira2011} is not able to appropriately correct the conduction band.

The corrected band structures of Fig.~\ref{bandstructures} present all the expected general features for the NV$^-$ center in diamond:
the defect $1a_1$ energy levels are resonant inside the valence band;
the relative positions of the spin up and spin down levels are correct; and
the first possible valence band excitation is high energetic enough, avoiding an electron transition from the valence band to the defect $2a_{1\downarrow}$ energy levels when the pumping laser is shined.

Although the $e_\downarrow$ energy levels appear to be closer to the conduction band than expected, due to the slightly underestimated band gap, the transition energies analysis is not impaired.
On the other hand, since the gap underestimation is a diamond particular case, as explained above, the method is expected to display still better performance when applied to other semiconductors, like the III-V ones, in which the application of the DFT-1/2 method presents very accurate results. \cite{Ferreira2011}

The usual procedure to obtain the optical transition energies via DFT is to take the difference between the total energy values of the excited and ground states.
The correction method proposed here allows these quantities to be extracted directly from the Kohn-Sham eigenvalues.
To verify this claim, Table~\ref{tableresults} presents results obtained with the usual total energy approach and with the eigenvalues approach, without the quasi-particle corrections.
Even though these values are not supposed to correspond to the quasi-particle band gap, they should agree with each other, and in fact, they do within a precision of 0.04~eV.

\begin{table}[t]
  \begin{threeparttable}
    \centering
    \caption{Vertical absorption (E$_{Ab}$), vertical emission (E$_{Em}$), zero-phonon line (E$_{ZPL}$), Stokes shift (E$_{S}$) and anti-Stokes shift (E$_{aS}$) energies calculated by different methods, compared to the experimental data~\cite{Davies1976} (all values in eV).}
    \label{tableresults}
    
    \begin{tabular}{*{6}{c}}
       \toprule
                                                       & E$_{Ab}$ & E$_{Em}$ & E$_{aS}$         & E$_{ZPL}$         & E$_{S}$             \\
       \midrule
        \mr{GGA}{(total energy)}                       & 1.90     & 1.55     & 0.16            & 1.71               & 0.19                \\
        \mr{GGA}{(eigenvalues)}                        & 1.86     & 1.55     & 0.16$^\ast$     & 1.72$^\dagger$     & 0.15$^\ddagger$     \\
        \mr{\bf{GGA-1/2}}{\bf{(eigenvalues)}}          &\bf{2.18} &\bf{1.68} &\bf{0.16}$^\ast$ &\bf{1.85}$^\dagger$ &\bf{0.33}$^\ddagger$ \\ 
        \mr{HSE06 \cite{Gali2009}}{(total energy)}     & 2.21     & 1.74     & 0.22            & 1.96               & 0.26                \\
        Exp.\cite{Davies1976}                          & 2.18     & 1.76     & 0.19            & 1.95               & 0.24                \\
       \bottomrule
    \end{tabular}
    \begin{tablenotes}
     \small
     \item $^\ast$ Calculated using Eq.~\ref{AS}
     \item $^\dagger$ Calculated using Eq.~\ref{ZPL}
     \item $^\ddagger$ Calculated using Eq.~\ref{SS}
    \end{tablenotes}
  \end{threeparttable}   
\end{table}

The results obtained when using the DFT-1/2 approximate quasi-particle corrections are in close agreement with the reported HSE results and experimental data.
It is observed that, in the GGA-1/2 results, the relative error of the Stokes and anti-Stokes shifts are greater than that of the other energies, as expected, since both shifts values result from the difference between two values very close to each other.

In our development, it has been argued that standard DFT approach may provide a good estimate of the anti-Stokes shift, and this is supported by the results from Ref.~\citenum{Gali2009} that reports both GGA-PBE and HSE calculations of the anti-Stokes shift for a larger supercell (4ax4ax4a) than the one used here, and they indeed shown that GGA slightly outperformed HSE.

\section{Conclusion}

The Kohn-Sham eigenvalues, with quasi-particle corrections, have been related to the experimental transition energies by introducing a new procedure, that allows the application  of the low computational effort method DFT-1/2 to correct the relative position of the deep defect energy levels. 
Since this method sharply reduces the computational cost when compared with other methods, it allows a systematic search for new defects in semiconductor hosts for several applications, including the search for new solid state qubits.
In particular, the NV$^-$ center in diamond has been considered as a benchmark of the method, and also as an example of its application.
In this test case, an accuracy of 0.1~eV have been reached in comparison with experimental data.


\begin{acknowledgments}
We thank the brazilian funding agencies FAPESP (grant n. 2012/50738-3), CAPES (PVE - grant n. 88887.116535/2016-00), and CNPq (grants n. 305405/2014-4, 308742/2016-8, and 154636/2016-9) for the financial support.
We acknowledge the National Laboratory for Scientific Computing (LNCC/MCTI, Brazil) for providing HPC resources of the SDumont supercomputer.
LVCA acknowledge partial support from CNPq (grant n. 312337/2013-2)

BL thanks professor Tom\'as Palacios for the introduction to the search for solid state qubits.
\end{acknowledgments}


\appendix

\section{Obtaining differences in total energy through eigenvalues}\label{eigenvalues}

Consider a situation in which we want to obtain the energy of an electronic transition between the localized states $\psi_\alpha$ and $\psi_\beta$ through the use of a supercell DFT calculation.
Defining the total energy $E$ in terms of partial occupations
\begin{align}
    \label{ETLD} &E = T + U[n] + E_{xc}[n],\\
    \label{NRHO} &n\big(\vec{r}\big) = \sum_i{f_i \big\lvert\psi_i\big(\vec{r}\big)\big\rvert^2},
\end{align}
where $n$ is the electron number density, $\psi_i$ is the \emph{i}-th Kohn-Sham orbital and $f_i$ its occupancy, $T$ is the kinetic energy, $U$ is the classical Coulomb energy, and $E_{xc}$ is the exchange-correlation functional.

Considering all but $\psi_\alpha$ and $\psi_\beta$ levels' occupations are fixed, we have $E$=$E(f_\alpha,f_\beta)$. The Janak's theorem states that
\begin{equation}
	\label{JANAK} \frac{\partial E}{\partial f_i}=\varepsilon_i,
\end{equation}
where $\varepsilon_i$ is the \emph{i}-th Kohn-Sham eigenvalue. In a large supercell, the excitation of a localized electron is a small perturbation on the Kohn-Sham operators. Thus, the position of the eigenvalues remain unchanged and we can immediately integrate to obtain
\begin{equation}
	\label{TED1} E(1,f_\beta) - E(0,f_\beta) = \varepsilon_\alpha, \forall f_\beta
\end{equation}
and
\begin{equation}
	\label{TED2} E(f_\alpha,1) - E(f_\alpha,0) = \varepsilon_\beta, \forall f_\alpha.
\end{equation}

We can express the transition energy of interest as
\begin{equation}
	\label{EIGEN} \Delta E_{trans} = E(0,1) - E(1,0) = \varepsilon_\beta - \varepsilon_\alpha,
\end{equation}
\emph{i.e.} the transition energy can be computed as the difference between the Kohn-Sham eigenvalues.


\bibliography{reflatex}

\end{document}